\documentclass[twocolumn,preprintnumbers,aps,prb,floatfix,showpacs,amsmath]{revtex4}

\usepackage[dvips]{color,graphicx}
\usepackage{epic,epsf}
\usepackage{dcolumn}
\usepackage{psboxit}


\begin{document}

\preprint{Phys.~Rev.~B {\bf 68}, 144519 (2003)}{}

\title{Electronic Collective Modes and Superconductivity in Layered
Conductors}

\author{A.~Bill}
\affiliation{Paul Scherrer Institute, Condensed Matter Theory, 5232
  Villigen PSI, Switzerland}
\author{H.~Morawitz}
\affiliation{IBM Almaden Research Center, 650 Harry Rd., San Jose, CA
95120, USA, and\\Institute for Theoretical Physics, Ulm University,
89069 Ulm, Germany}
\author{V.Z.~Kresin}
\affiliation{Lawrence Berkeley Laboratory, University of California,
Berkeley, CA 94720, USA}

\date{\today}

\begin{abstract}
A distinctive feature of layered conductors is the presence of
low-energy electronic collective modes of the conduction
electrons. This affects the {\it dynamic} screening properties of the
Coulomb interaction in a layered material. We study the consequences
of the existence of these collective modes for
superconductivity. General equations for the
superconducting order parameter are derived within
the strong-coupling phonon-plasmon scheme that account for the screened
Coulomb interaction. Specifically, we calculate the superconducting
critical temperature $T_c$ taking into account the full temperature,
frequency and wave-vector dependence of the dielectric function. We
show that low-energy plasmons may contribute constructively to
superconductivity. Three classes of layered superconductors are
discussed within our model: metal-intercalated halide nitrides,
layered organic materials and high-$T_c$ oxides. In particular, we
demonstrate that the plasmon contribution (electronic mechanism) is
dominant in the first class of layered materials. The theory shows
that the description of so-called ``quasi-two-dimensional
superconductors'' cannot be reduced to a purely 2D model, as commonly
assumed. While the transport properties are strongly anisotropic, it
remains essential to take into account the screened {\it inter}layer
Coulomb interaction to describe the superconducting state of layered
materials.
\end{abstract}


\maketitle

\section{Introduction}\label{s:intro}

Recent years have witnessed the discovery of many new superconducting
materials: high-temperature cuprates, fullerides, borocarbides,
ruthenates, MgB$_2$, metal-intercalated halide nitrides,
intercalated Na$_x$CoO$_2$\cite{takada03}, etc. Systems such as
organics, heavy
fermions, nanoparticles have also been intensively studied. Many of
these systems belong to the family of layered conductors
characterized, e.g., by strongly anisotropic electronic transport
properties. Recently, it was reported that even Ba$_{1-x}$K$_x$BiO$_3$
has a layered structure.\cite{klinkova03} An interesting question
raised by the observation of superconductivity in all the systems
mentioned above is the following: why is layering a favorable factor
for superconductivity? The present paper addresses this
question. We show that layering leads to peculiar dynamic screening of
the Coulomb interaction and that this is important for the description of
the superconducting state in layered conductors.

The conventional theory of superconductivity has mostly dealt with
three-dimensional (3D) isotropic systems, although some papers have also
described the impact of the Fermi surface anisotropy on the
superconducting state (see, e.g., review Ref.~\onlinecite{clem}). In
this theory the Coulomb repulsion is described by a static
pseudopotential $\mu^{\star}$ and its value is reduced because of the
well-known logarithmic factor $\ln(E/\Omega)$ where $E$ is an
electronic energy and $\Omega$ is a characteristic bosonic
(e.g.~phonon) energy. Such a static approach is justified by the large
value of the plasmon frequency $\Omega_{pl}({\bf q}=0) =
\min\{\Omega({\bf q})\} \equiv \Omega_{pl}$ in usual metals, where
$\Omega_{pl}$ ranges between 5eV and 30eV. Such high energies imply a
perfect, instantaneous screening of the Coulomb interaction.

Layered conductors have a structure of the plasmon spectrum that
differs fundamentally from 3D metals. In addition to the high energy
``optical'' collective mode mentioned above, the spectrum contains
also an important low-frequency part (see below). The screening of the
Coulomb interaction is incomplete and the {\it dynamic} nature of the
interaction becomes important. As a result, the interplay between the
attractive interaction and the Coulomb term is
more subtle than introduced in the conventional theory of
superconductivity. It is on this screened Coulomb term and its
interplay with the electron-phonon mechanism that we focus in the
present paper.

Our goal is to evaluate the additional impact of dynamic screening
on pairing in layered superconductors. The pure plasmon mechanism
(that is, in absence of any other attractive interaction) has been
discussed previously for 3D and 2D systems (see, e.g.,
Refs.~\onlinecite{3Dplasmons,vertex,ruvalds,takada1})
The acoustic plasmons for spatially separated layers in
  metal-oxide semiconductor structures were introduced and analyzed by
  Y.~Takada in Ref.~\onlinecite{takada2}. The author indicated the
  possibility of acoustic-plasmon mediated superconductivity.
In the present paper we focus on {\it layered} conductors.
More importantly, we consider plasmons' contribution {\it in
  conjunction} with the phonon mechanism. It is assumed that the
phonons themselves provide the pairing
  so that at $T=0$K the compound is in the superconducting
  state. In other words, the presence of phonons is sufficient to
  overcome the static Coulomb repulsive interaction. Within this
  scenario the dynamic screening acts as an additional
  factor. Therefore, in the absence of the plasmon term we obtain
  the conventional Eliashberg equations; the electron-phonon
  coupling constant and the Coulomb pseudopotential are thus
  considered as parameters to be determined from experimental data
  (see, e.g., Ref.~\onlinecite{mcmillanwolf}). Note that the contribution
of phonons and plasmons to the superconducting state has aslo been considered
in Ref.~\onlinecite{takada5} for fullerides. We also point out that we
consider the electron-phonon interaction (phonon-plasmon mechanism) for
concreteness. However, since our attention is set on the Coulomb
contribution to the total pairing, our approach is valid for other
mechanisms as well. The advantage of the present approach is that we
are not restricted to answer the question whether or not plasmons
themselves can provide superconductivity, but allows to answer the
question whether low-energy plasmons can sustain or enhance the
pairing induced by other mechanisms.

We discussed briefly our approach in
Refs.~\onlinecite{kresin87,kresin90,morawitz93,bill99}. The present
article contains a
detailed analysis of the dielectric function, the plasmon spectrum and
its impact on superconductivity in layered
superconductors. Furthermore, we apply the theory to characteristic
examples of three classes of materials: the metal-intercalated halide
nitrides, organic and high-temperature superconductors.

The structure of the paper is as follows. In section \ref{s:maineqs}
we present the main equations describing layered superconductors and
discuss the electron-phonon and Coulomb contributions to the
pairing-interaction kernel. In section \ref{s:collmodes} we discuss
the dynamic screening of the Coulomb interaction in layered
conductors. The dielectric function and the resulting
electronic collective excitations (layer plasmons) will be
described. It is essential that the dielectric function is evaluated
and analyzed in the thermodynamic Green's function formalism; this
allows us to calculate $T_c$. In the next section,
Sec.~\ref{s:LC}, we consider three classes of
layered superconductors: the metal-intercalated halide nitrides, the
organics and high-$T_c$ superconductors. The conclusions are presented
in Sec.~\ref{s:conclusions}.

\section{Main Equations}\label{s:maineqs}

\begin{center}
\begin{figure}
\vspace*{.5cm}
\includegraphics[width=6.5cm]{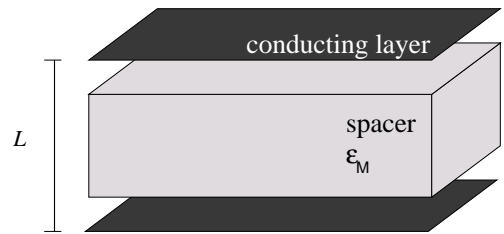}
\caption{\label{Fig:model}
The layered electron gas (LEG) model. The conducting sheets (dark) are
stacked along $c$ and separated by spacers (light) with dielectric
constant $\epsilon_M$. The model considers an infinite stacking of
layers. The electrons are moving within the conducting sheets. The
Coulomb interaction is effective within, but also between the sheets
(see text). $L$ is the interlayer distance.}
\end{figure}
\end{center}

We consider a layered system consisting of stacks of conducting sheets
along the $z$-axis separated by dielectric spacers (see
Fig.~\ref{Fig:model}). Because of the conductivity's high
anisotropy it is a good approximation to neglect transport between the
layers (see Sec.~\ref{s:LC}). On the other
hand, the Coulomb interaction between charge-carriers is effective
both {\it within} and {\it between} the sheets. To ensure
charge-neutrality we further introduce positive counter-charges spread
out homogeneously over the sheets.
\unitlength1cm
In order to calculate the critical temperature $T_c$ for the
superconducting transition we start with the equation for the
superconducting order parameter $\Delta({\bf k},\omega_n)$ and the
renormalization function $Z({\bf k},\omega_n)$:
\begin{subequations}
\label{DeltaZ}
\begin{eqnarray}
\label{Delta}
\Delta({\bf k},\omega_n) Z({\bf k},\omega_n)
=&&\\
&&\hspace*{-1.5cm}T \sum_{m=-\infty}^{\infty} \int \frac{{\rm d}^3{\bf k'}}{(2\pi)^3}
\Gamma({\bf k}, {\bf k'}; \omega_n-\omega_m)
F^{\dagger}({\bf k'},\omega_m),\nonumber\\
\label{Z}
Z({\bf k},\omega_n) - 1
=&&\\
&&\hspace*{-1.5cm}\frac{T}{\omega_n} \sum_{m=-\infty}^{\infty} \int \frac{{\rm d}^3{\bf
    k'}}{(2\pi)^3} \Gamma({\bf k}, {\bf k'}; \omega_n-\omega_m)
G({\bf k'},\omega_m),\nonumber
\end{eqnarray}
\end{subequations}
where
$F^{\dagger} = <c^{\dagger}_{{\bf k},\uparrow}c^{\dagger}_{-{\bf
    k},\downarrow}>$ is the Gor'kov pairing function,
$G = <c^{\dagger}_{{\bf k},\sigma}c_{{\bf k},\sigma}>$ is the usual
Green function, and $\Gamma$ is the total interaction kernel.

These equations can be rewritten in the following form (at $T=T_c$)
\begin{widetext}
\begin{subequations}
\label{phiZ0}
\begin{eqnarray}
\label{phi0}
\phi_n({\bf k}) &=&
T \sum_{m=-\infty}^{\infty} \int \frac{{\rm d}^3{\bf k'}}{(2\pi)^3}
\Gamma({\bf k}, {\bf k'}; \omega_n-\omega_m)
\frac{\phi_m({\bf k'})}{\omega^2_m({\bf k'}) + \xi_{\bf k'}^2}
\Bigg|_{T_c},\\
\label{Z0}
\omega_n({\bf k}) - \omega_n &=&
T \sum_{m=-\infty}^{\infty}\int \frac{{\rm d}^3{\bf k'}}{(2\pi)^3}
\Gamma({\bf k}, {\bf k'}; \omega_n-\omega_m)
\frac{\omega_{m}({\bf k'})}{\omega^2_m({\bf k'}) +
  \xi_{\bf k'}^2} \Bigg|_{T_c}.
\end{eqnarray}
\end{subequations}
\end{widetext}
In these equations and in the rest of the paper we use the following
notations: ${\bf k} = ({\bf k_{\parallel}}, k_z)$, where the $z$-axis
is chosen to be perpendicular to the layers. We use the
thermodynamic Green's function formalism (see, e.g.,
Ref.~\onlinecite{mahan}) with $\omega_n = (2n+1)\pi T$. Because of the
relation $\omega_n-\omega_m = 2\pi T (n-m)$ we often
use the short hand ($n-m$) to denote the frequency dependence [e.g.,
($n+m+1$) stands for $\omega_n-\omega_{-(m+1)}$]. Finally, we define
$\phi_n({\bf k})\equiv\Delta_n({\bf k}) Z_n({\bf k})$,
$\omega_n({\bf k}) \equiv\omega_{n}Z_n({\bf k})$, $\Delta_n \equiv
\Delta(\omega_n)$ and $Z_n \equiv Z(\omega_n)$.

For concreteness we focus on the case where the interaction kernel is
a sum of electron-phonon and Coulomb interactions.
Then, the total kernel $\Gamma\equiv \Gamma({\bf q},\omega_n-\omega_m)$,
with ${\bf q}={\bf k}-{\bf k'}$, is written in the form
\begin{equation}\label{gamma}
\Gamma = \Gamma_{ph} + \Gamma_c,
\end{equation}
with
\begin{eqnarray}
\label{gammaph}
\Gamma_{ph}({\bf q};|n-m|) &=&
|g_{\nu}({\bf q})|^2 D({\bf q},|n-m|) \nonumber\\
&=&
|g_{\nu}({\bf q})|^2 \frac{\Omega_{\nu}^2({\bf
    q})}{(\omega_n-\omega_m)^2 + \Omega_{\nu}^2({\bf q})},\\
\label{gammac}
\Gamma_c ({\bf q};|n-m|) &=& \frac{V_c({\bf q})}{\epsilon({\bf q},|n-m|)}.
\end{eqnarray}
$D({\bf q},n-m)$ is the phonon Green function,
$\Omega_{\nu}^2({\bf q})$ the phonon dispersion relation; summation
over phonon branches $\nu$ is assumed. The second, important Coulomb term
$\Gamma_c$ is written in its most general form as the ratio of the
bare Coulomb interaction $V_c({\bf q})$ and the dielectric function
$\epsilon({\bf q},\omega_n-\omega_m)$. Both the Coulomb interaction
and the dielectric function have to be calculated for a layered
structure.
It should be noted that the relation between these two interactions
[$\Gamma_{ph}$ and $\Gamma_c$, Eq.~(\ref{gamma})] is more subtle. For
example, the Coulomb screening affects the value of the
electron-phonon matrix elements (see, e.g.,
Ref.~\onlinecite{geilikman75}). Here, however, we do not calculate the
electron-phonon coupling constant $\lambda$ [see Eq.~(\ref{lph})] and,
similarly to the treatment of conventional superconductors (see, e.g.,
Ref.~\onlinecite{mcmillanwolf}), we use the values determined from
experimental data. For example, in the case halide nitrides considered
in Sec.~\ref{ss:nitrides} $\lambda$ was determined from heat capacity
measurements \cite{tou01}.

The Coulomb potential $V_c({\bf q})$ is the Fourier
transform of the 3D Coulomb interaction $V_c({\bf r}) = e^2/\epsilon_M
|{\bf r}|$, where $\epsilon_M$ is the dielectric constant of the
spacers, and takes the form (see appendix \ref{a:Coulomb})
\begin{eqnarray}
\label{Vcq}
V_c({\bf q}) = \frac{2\pi
e^2}{\epsilon_Mq_{\parallel}}R(q_{\parallel},q_z),
\end{eqnarray}
where $R(q_{\parallel},q_z)$ is defined in Eq.~(\ref{VcR}) below.
Introducing dimensionless quantities $\tilde{q} = q_{\parallel}/2k_F$,
$\kappa_F=2k_FL$ ($L$ is the interlayer distance, $k_F$ the in-plane
Fermi wave-vector) as well as $N(0) = m_b/2\pi\hbar^2$, the 2D
electronic density of states ($m_b$ is the band mass), we can write
\begin{eqnarray}
\label{Vcqul}
V_c({\bf q}) = \frac{\lambda_c}{N(0)} \frac{R(\tilde{q};q_z)}{\tilde{q}},
\end{eqnarray}
with
\begin{eqnarray}
\label{VcR}
R(\tilde{q},q_z) &=& \frac{\sinh(\kappa_F\tilde{q})}{\cosh(\kappa_F\tilde{q}) -
\cos(q_zL)}(1-\delta_{{\bf q},0}).
\end{eqnarray}
Eq.~(\ref{Vcq}) contains the product of the two-dimensional Coulomb
potential and the function $R(q_{\parallel},q_z)$ which reflects the
layered nature of the studied system. As expected, $\lim_{L\to \infty}
R({\bf q}) = 1$, whereas $R({\bf q}) = 2 q_{\parallel}/|{\bf q}|^2L$ for
$L \ll 1$. Furthermore, $V_c$ diverges as $1/|{\bf q}|^2$ for $|{\bf
  q}|\to 0$, in agreement with the 3D character of this limit. Note
that $R(\tilde{q};q_z)$ contains the factor $(1 - \delta_{{\bf q},0})$
reflecting the presence of the neutralizing positive ion
counter-charges; the presence of this term is implicit in the
following.

Eq.~(\ref{Vcqul}) contains the dimensionless Coulomb interaction
constant defined by
\begin{eqnarray}\label{lamc}
\lambda_c &=& \frac{1}{2\epsilon_M}\left(\frac{e^2}{\hbar v_F}\right)
=
\frac{r_s}{\sqrt{8}}
=
\frac{\alpha}{2}\frac{c}{v_F}.
\end{eqnarray}
$v_F$ is the Fermi velocity, $c$ the vacuum speed of light and
$\alpha$ is the fine structure constant. Note that $\lambda_c\sim r_s$,
where $r_s=\sqrt{2}/k_Fr_B$ ($r_B=\hbar^2\epsilon_M/m e^2$ is the Bohr radius)
is the well-known dimensionless electron density radius defined here for a
layered electron gas.

The electronic screening of the Coulomb interaction is
described by the dielectric function $\epsilon({\bf q},\omega_n)$
written in its most general form as
\begin{eqnarray}\label{dielLEG}
\epsilon({\bf q},\omega_n-\omega_m)
= 1 - V_c({\bf q})\Pi({\bf q},\omega_n - \omega_m).
\end{eqnarray}
In the following we use the RPA method. As is known (see, e.g.,
  Ref.~\onlinecite{nozieres}) for real 3D metals the RPA provides a
  qualitative description, whereas a quantitative analysis requires to
  go beyond this approximation. For the systems of interest RPA is
  favorable because of the inequality $\lambda_c < 1$ (see
  below). Note that the contribution of the background dielectric
  function and the inequality $\epsilon_M > 1$ could be essential
  (cf.~e.g.~Ref.~\onlinecite{geilikman68}). It would be interesting
  to perform more exact calculations using methods as those of
  Ref.~\onlinecite{takada3} and, in addition, take into account the
  band structure of real materials instead of the LEG model. We think
  that the approximation based on the inequality $\lambda_c<1$
  provides the adequate physical picture.

\section{Layered conductors: Electronic Collective Modes}
\label{s:collmodes}

\subsection{Plasmon bands}\label{ss:plasmons}

The spectrum of collective electronic excitations is determined by
the poles of the two-particle Green function which coincides with the
poles of the vertex $\Gamma_c({\bf q},\omega)$. The latter is the
analytic continuation (see, e.g., Ref.~\onlinecite{mahan}) of the
function $\Gamma_c({\bf q},\omega_n)$, Eq.~(\ref{gammac}). These poles
correspond to the zeros of the real-frequency dielectric function
$1-V_c({\bf q})\Pi({\bf q}_{\parallel},\omega)=0$. At $T=0$ the real
part of the polarizability of a single layer takes the form
($\omega>\hbar q_\parallel v_F$)\cite{kresin90,morawitz93}
\begin{eqnarray}
\label{RealPi}
Re\left\{\Pi({\bf q},\omega)\right\} =
2 N(0)\left[ \frac{\omega}{\sqrt{\omega^2 - (\hbar q_\parallel
      v_F)^2}} - 1\right] .
\end{eqnarray}
For $\omega\gg \hbar q_\parallel v_F$ this expression reduces to
$Re\left\{\Pi\right\} \simeq N(0)\hbar^2q_\parallel^2v_F^2/\omega^2$,
as obtained in Ref.~\onlinecite{morawitz93}.

From Eqs.~(\ref{Vcqul}), (\ref{VcR}) and (\ref{RealPi}) we
derive the general expression for the plasmon dispersion relation:
\begin{eqnarray}
\label{pldisp}
\omega = \hbar q_\parallel v_F \sqrt{1 +
  \frac{(N(0)V_c)^2}{\frac{1}{4} + N(0)V_c}},
\end{eqnarray}
where $V_c\equiv V_c(\tilde{q}, q_z)$ is the Coulomb interaction
defined in Eq.~(\ref{Vcqul}). If $N(0)V_c\gg 1$ we obtain the optical
plasmon $\omega = \hbar q_\parallel v_F \sqrt{1+N(0)V_c}$ 
(this corresponds to the hydrodynamic approximation for small
${\bf q}$; see Ref.~\onlinecite{fetter}). 
The plasmon band $\omega = \omega(q_\parallel,q_z)$ is confined
between the upper branch with $q_z=0$ (in-phase motion of the charge
carriers) and the lower branch at $q_z=\pi/L$ (out-of-phase motion of
carriers). Indeed, for $\omega\gg \hbar q_\parallel v_F$
Eq.~(\ref{pldisp}) reduces to the expression $\omega \simeq
\hbar q_\parallel v_F \sqrt{N(0)V_c}$ which at $q_z=0$
leads to the usual ``optical'' plasmon with $\Omega_{pl}^2 \equiv
\omega^2(q_\parallel=0,q_z=0) = 4e^2\varepsilon_F/\epsilon_M L$. For
$q_z=\pi/L$, on the other
hand, we obtain the dispersion law for the ``acoustic'' plasmon
(linear in $q_\parallel$) of the form of Eq.(\ref{pldisp}) with
\begin{eqnarray}
N(0)V_c\left(\tilde{q},q_z=\frac{\pi}{L}\right) =
\frac{\lambda_c}{\tilde{q}}
\frac{\sinh(\kappa_F\tilde{q})}{\cosh(\kappa_F\tilde{q}) + 1}
\stackrel{q_\parallel \to 0}{\longrightarrow} \lambda_c k_F L.
\end{eqnarray}
For $q_z=\pi/L$ and $q_\parallel L \ll 1$, we obtain $\omega \approx
(\Omega_{pl} L/2) q_\parallel$.

\begin{center}
\begin{figure}[htp]
\includegraphics[width=6.5cm]{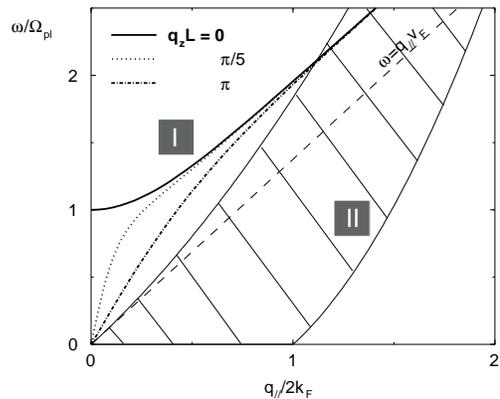}
\caption{\label{Fig:pldisp}
Electronic excitation spectrum for the
  layered electron gas (LEG) (see also
  Ref.~\onlinecite{morawitz93}). The solid, dotted and
  dash-dotted lines in region I ($\omega>\hbar q_\parallel v_F$) are
  plasmon dispersion relations corresponding to the $q_z$ indicated on
  the figure. The area delimited by the $q_z=0$ (solid) and $q_z =
  \pi/L$ (dash-dotted) bands
  contains the dispersion relations for all $q_z$. The branch $q_z =
  \pi/5L$ (dotted) is shown as an example. The hashed area denotes the
  electron-hole excitation continuum in which plasmon (Landau) damping
  occurs. The long-dashed line separates region I ($\omega>\hbar
  q_\parallel v_F$) from region II ($\omega<\hbar q_\parallel v_F$).}
\end{figure}
\end{center}

Thus, the plasmon spectrum of a layered conductor, Eq.~(\ref{pldisp}),
has the rather complicated structure shown in Fig.~\ref{Fig:pldisp}.
\unitlength1cm

The dispersion can be viewed as a continuous set of
``acoustic'' branches parametrized by $q_z \in [0,\pi/L]$ (the slope
of the acoustic plasmon at $q_{\parallel}\to 0$ is smallest for
$q_z=\pi/L$ and increases as $q_z\to 0$).
Only the upper branch for $q_z=0$ represents an ``optical'' branch
and, as expected for the long wavelength limit, corresponds to the
usual 3D plasmon.
Crucial for the phenomenon of dynamic screening and its
impact on the pairing is the presence of the low-energy collective
excitations, which can play a role similar to phonons (they can be
labeled ``electronic'' sound).

Note that the low-energy plasmon branches, so-called ``demons'', also
appear in the presence of different overlapping bands (e.g.~``light''
and ``heavy'' carriers; see, Ref.~\onlinecite{3Dplasmons} and the
review Ref.~\onlinecite{ruvalds}). We emphasize that the case
considered in the present paper is entirely different. Indeed, the
appearance of ``acoustic'' branches is caused by the presence of
spatially separated conducting layers and the out-of-phase motion of
the carriers in neighbouring planes.

The partial density of states can be determined for each plasmon band
(corresponding to each $q_z$) from the dispersion relation,
Eq.~(\ref{pldisp}). As first pointed out by two of the authors,
Bozovic {\it et al.} in Ref.~\onlinecite{morawitz93}, the density of
states considered as a function of energy is peaked
at the boundaries, that is for $q_z=0$ and $q_z=\pi/L$. A good
approximation is thus to model the plasmon spectrum of a layered
conductor as consisting of two branches: the upper ``optical'' branch
($q_z=0$) and the lower ``acoustic'' branch ($q_z = \pi/L$). We have
shown earlier\cite{bill99} that the optical 
branch gives an essentially repulsive contribution to the pairing
interaction. We therefore include this latter part in the effective
repulsive $\mu^\star$ of region II (Fig.~\ref{Fig:pldisp}). In the
following we consider only the contribution of the dominant acoustic
branch at $q_z=\pi/L$. The contribution of all other branches only
enhances the effect of the $q_z=\pi/L$ branch, as we discuss below. It
is worth emphasizing that the existence of the latter branches is
specific to layered materials.

We end this section by noting that the inclusion of a residual
interlayer hopping would imply the appearence of a small gap for the
acoustic plasmons. The size of this gap is determined by the
interplane hopping parameter $t_z$. The more isotropic the system becomes,
the larger is the gap. In the isotropic limit the one
degenerate optical plasmon branch observed in 3D metals is recovered. As 
mentioned in the introduction, however, the materials of interest for
the present paper (Sec.~\ref{s:LC}) have a ratio $t_z/t_\parallel\lesssim
10^{-3}$ so that discarding interlayer transport is a good
approximation. Further support for this approximation
  is found in Ref.~\onlinecite{falter02a,falter02b} (and references
  therein) from dielectric properties and lattice dynamics studies of
  high-temperature superconductors.

\subsection{Screening of the Coulomb interaction: the dielectric
  function at finite temperature.}\label{ss:poldiel}

To study the impact of dynamic screening on the superconducting
state we need to calculate the dielectric function,
Eq.~(\ref{dielLEG}), which contains the polarizability $\Pi({\bf
  q},\omega_n)$. In particular, to obtain $T_c$, we have to determine
these functions at {\it finite} temperatures. In RPA the
polarizability takes the well-known form
\begin{equation}\label{pol}
\Pi({\bf q}_{\parallel},i\omega_n) = 2\int d^2{\bf k}_{\parallel}\,
\frac{f_{{\bf k}_{\parallel}} - f_{{\bf k}_\parallel + {\bf
      q}_{\parallel}}}{i\omega_n + \xi_{{\bf k}_{\parallel}} -
  \xi_{{\bf k}_\parallel + {\bf q}_{\parallel}}}.
\end{equation}
where $f_k \equiv f(\xi_{{\bf k}_{\parallel}})$ is the Fermi
distribution and all wave vectors lie in the plane of the layered
structure. To the 
best of our knowledge all previous works concerned with
layered structures were done either using the calculated
polarizability at $T=0$ (see, e.g., Ref.~\onlinecite{stern}) or taking
the static limit for non-zero temperatures (as done, e.g., in
Ref.~\onlinecite{3Dplasmons,vertex}) Here we calculate the
polarizability both at finite temperatures (using the temperature
Green's function formalism) {\it and all} values of ({\bf
  q},$\omega_n$).
As the calculations of the next sections will show, the
  temperature dependence of the polarizability can be neglected in
  some cases (e.g.~for halide nitrides,
  Sec.~\ref{ss:nitrides}) but should be taken into account for the
  cuprates where the ration $T_c/\varepsilon_F$ is not negligibly small
  (see Sec.~\ref{ss:HTSC}). In general, the proper account
  of dynamic screening requires to consider all three parameters.
Note that to render the numerical problem tractable when solving
Eq.~(\ref{DeltaZ0xi}) below, we reduce the number of integrals to be
performed numerically by writing Eq.~(\ref{pol}) in polar coordinates
and integrating analytically over the angles (cf.~appendix
\ref{a:pol}). The remaining $k$-integration is then done numerically.

\begin{center}
\begin{figure}[htb]
\includegraphics[width=6.5cm]{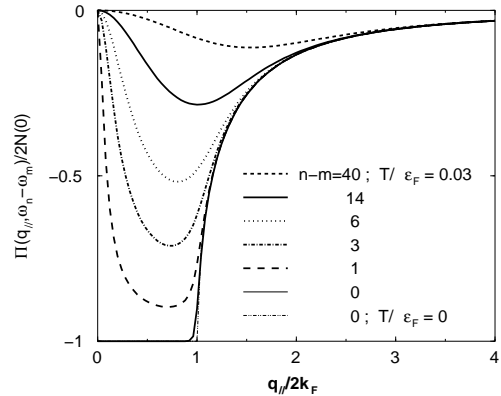}
\caption{\label{Fig:polLEG}
Electronic polarizability in RPA as a function of
$\tilde{q}=q_\parallel/2k_F$ for various values of $\omega_n-\omega_m
= 2(n-m)\pi T$ (from top to bottom) and a typical temperature
$T/\varepsilon_F=0.03$. The lowest, dash-dotted curve is obtained for
$T=0$K and remains the same for {\it all} frequencies [i.e.~at $T=0$,
$\Pi(q<2k_F,\omega_n-\omega_m) = -2N(0)$].}
\end{figure}
\end{center}

Fig.~\ref{Fig:polLEG} displays the polarizability, Eq.~(\ref{pol}), of
the electron gas of a layer, as a function of wave-vector
$\tilde{q}=q_\parallel/2k_F$ for different values of the frequency
$\omega_n$ and a typical temperature $T/\varepsilon_F=0.03$ which applies
to high-$T_c$ superconductors (see Sec.~\ref{ss:HTSC}). We first point
out that the $\tilde{q}$-dependence of this function is essentially
restricted to the interval $[0,4k_F]$.
\unitlength1cm

Let us now consider the temperature dependence of the
polarizability.
First we discuss the case $T=0$. For the real-frequency
polarizability the analytical continuation gives the result derived by
Stern in Ref.~\onlinecite{stern}. On the other hand, in the Matsubara
temperature-Green's function formalism the polarizability does not
depend on frequency for $\omega < q_\parallel v_F$
(cf.~Ref.~\onlinecite{mahan}a, Sec.~20.2):
$\Pi(q,\omega_n) =
-2N(0)= - m_b/\pi \hbar^2$ (bottom dotted line of
Fig.~\ref{Fig:polLEG}).
 For $T>0$ the polarizability vanishes at high frequencies as shown in
Fig.~\ref{Fig:polLEG}. The higher the temperature, the smaller is the 
frequency-range over which the polarizability remains finite. Note
that the shape of $\Pi(\tilde{q},\omega_n-\omega_m=0)$ (lower solid line
on the figure) is almost unaltered until very high temperatures. This
can also be directly seen from the analytical expression
[see appendix, Eq.~(\ref{polq01})].

\unitlength1cm
\begin{center}
\begin{figure}[h]
\includegraphics[width=6.5cm]{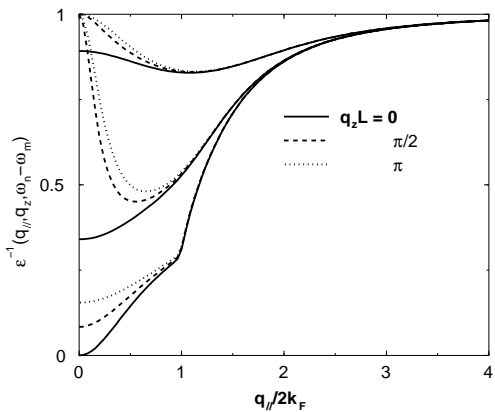}
\caption{\label{Fig:dielLEG}
Inverse dielectric function of the LEG. The solid, dashed, and dotted
lines correspond to $q_zL=0,\pi/2,\pi$ respectively. The three bottom
curves are obtained for $\omega_n-\omega_m = 0$, the three middle
curves are for $n-m=10$ and the three uppermost curves are for
$n-m=40$. The other parameters were given in the previous figure. Note
that $\epsilon^{-1}({\bf q}\to 0,\omega_n-\omega_m)$ is zero (perfect
screening) only in the static limit.}
\end{figure}
\end{center}
Using the above results for the polarizability we calculate the
dynamic dielectric function, Eq.~(\ref{dielLEG}), or rather it inverse
$\epsilon^{-1}({\bf q},\omega_n-\omega_m)$ since it is this quantity that
enters the vertex $\Gamma_c$, Eq.~(\ref{gammac}). The result is shown 
on Fig.~\ref{Fig:dielLEG} for the same values of parameters as in
Fig.~\ref{Fig:polLEG}.
We point out a few important properties of the inverse dielectric
function. This function is bounded for all {\bf q}
and $\omega_n-\omega_m$. For high frequencies and/or large wave-vectors
$\epsilon^{-1}({\bf q},\omega_n-\omega_m)\to 1$, meaning that the Coulomb
interaction is unscreened in these cases. In fact, it can be shown
that for $\hbar q v_F \ll \omega_n-\omega_m$ the dielectric function
takes the form\cite{nozieres}
\begin{eqnarray}\label{diel0}
\epsilon(\hbar q v_F \ll \omega_n-\omega_m) \simeq \epsilon_D(n-m),
\end{eqnarray}
where
\begin{eqnarray}\label{diel0b}
\epsilon_D(n-m) \equiv 
1 + \frac{\Omega^2_{pl}}{(\omega_n-\omega_m)^2}.
\end{eqnarray}
Eq.~(\ref{diel0}) describes the dielectric function in the Drude limit.
Note that latter expression is exact in the limit ${\bf q}=0$ and
$\omega>0$.\cite{nozieres}

The result of Fig.~\ref{Fig:dielLEG} shows that for {\it any finite}
frequency, the long wave-length limit takes the form $\epsilon(|{\bf
  q}|\to 0,\omega_n-\omega_m>0)\to
\epsilon_D(\omega_n-\omega_m)$. Thus, it is only in the static case
$\omega_n-\omega_m=0$ that the Coulomb interaction is ``perfectly''
screened (exponential screening in real space): $\epsilon({\bf q}\to
0,\omega_n-\omega_m=0) \approx [1 + \kappa_{TF}^2/|{\bf q}|^2]^{-1}$. The
latter limit is the so-called Thomas-Fermi screening of the Coulomb
potential. For the LEG the screening length is given by
$\kappa_{TF}^2 = \pi N(0)\Omega_{pl}^2$. In all other cases the limit
of long wave-lengths is given by the Drude limit.

The dielectric function describing the screening in layered conductors
will be used in the next section to calculate the effect of the
dynamically screened Coulomb interaction on $T_c$ in several classes
of layered superconductors. We thereby use the full wave-vector,
frequency and temperature dependence of $\epsilon({\bf
  q},\omega_n-\omega_m)$ calculated in this section.

\section{Application to various layered systems}\label{s:LC}

In this section we consider the phonon-plasmon mechanism and in
particular the impact of the dynamic screening of the Coulomb
interaction on the superconducting state of several layered
systems. We discuss specific examples belonging to three classes of
materials: metallochloronitrides, organics and high-$T_c$
superconductors. To this aim, we first rewrite
Eqs.~(\ref{phi0},\ref{Z0}) in a form adequate for layered conductors
and convenient for calculations. We then evaluate the critical
temperature $T_c$ of the various compounds.

\subsection{Numerical analysis}\label{ss:numerics}

We assume isotropy within the layers. The order parameter and the
renormalization function can therefore be written as
$\Delta_n({\bf k})\simeq \Delta_n({\bf k}_{\parallel}={\bf k}_F,k_z)
\equiv \Delta_n(k_z)$ and $Z_n({\bf k}) \equiv Z_n(k_z)$ (see appendix
\ref{a:integ}). This approximation is valid since the Cooper
instability (see, e.g., Ref.~\onlinecite{mahan})
and, correspondingly, the pairing, occurs on the Fermi surface. As for
the integrands in Eqs.~(\ref{DeltaZ}-\ref{gammac}), they depend mainly on
the momentum transfer ${\bf q}_\parallel$. Thus, in
the layered electron gas the order parameter depends on the frequency
and the wave-vector perpendicular to the layers.

To perform numerical calculations with Eqs.~(\ref{phi0},\ref{Z0}) we
follow the standard procedure adapted to the case of layered
materials. We first express the integral over ${\bf k}_\parallel$ in
terms of an integral over energy and in-plane wave-vector amplitude
and carry out the former analytically (see appendix \ref{a:integ}). We
thereby reduce the equations to a form containing one-dimensional {\bf
  k}-space integrals. Note, however, that the dielectric function also
contains an integral to be performed at each iteration of the
calculation [Eqs.~(\ref{dielLEG},\ref{pol})]. The resulting equations
take the form (appendix \ref{a:integ})
\begin{subequations}
\label{DeltaZ0xi}
\begin{eqnarray}
\label{Delta0xi}
\phi_n(k_z)=\hspace*{.5cm}&&\\
&&\hspace*{-1.5cm}
\pi T \sum_{m=-\infty}^{\infty}
\frac{1}{N_z} \sum_{k_z'=-\pi}^\pi \bar{\Gamma}(q_z,n-m)
\frac{\phi_m(k_z')}{|\omega_m(k_z')|},\nonumber\\
\label{Z0xi}
\omega_n(k_z) - \omega_n =\hspace*{.5cm}&&\\
&&\hspace*{-1.5cm}
\pi T \sum_{m=-\infty}^{\infty}
\frac{1}{N_z} \sum_{k_z'=-\pi}^\pi \bar{\Gamma}(q_z,n-m)
\frac{\omega_m}{|\omega_m|},\nonumber
\end{eqnarray}
\end{subequations}
where $N_z$ are the number of $k_z$ points taken in the first Brillouin
zone and the kernel is given by
\begin{eqnarray}
\label{gammabar}
\bar{\Gamma}(q_z,n-m) 
&=&\lambda D(n-m) + \lambda_c \Gamma_c(q_z,n-m),
\end{eqnarray}
with
\begin{subequations}
\label{gammabarparts}
\begin{eqnarray}
\label{lph}
\lambda \equiv
\frac{N(0)}{\pi}\,\int_0^1\,
\frac{\rm{d}\tilde{q}}{\sqrt{1-\tilde{q}^2}}\,|g_\nu(\tilde{q})|^2,\hspace*{1cm}\\
\label{gammactilde}
\Gamma_c(q_z,n-m) \equiv\hspace*{5cm}&&\\
\hspace*{3.5cm}\frac{1}{\pi} \int_0^1 \frac{\rm{d}\tilde{q}}{\sqrt{1-\tilde{q}^2}}\,\,
\frac{R(\tilde{q},q_z)/\tilde{q}}{\epsilon(\tilde{q},q_z,n-m)}.\nonumber
\end{eqnarray}
\end{subequations}
All other quantities were defined in Sec.~\ref{s:maineqs}. Two
simplifications are made in the following calculations that allow to
single out the effect of low-energy electronic collective modes on
superconductivity. The first is to replace the phonon contribution to
the pairing to one (or two, see below Sec.~\ref{ss:organics})
characteristic phonon modes. The second is that we can set
$k_z=\pi/L$, based on the analysis made in
Sec.~\ref{ss:plasmons}, which shows that this wave-vector gives the
largest contribution to the pairing. Consequently, the order parameter
and the renormalization function taken at the zone boundary along
$k_z$ are function of frequency $\omega_n$ only. We emphasize,
however, that these two simplifications are not affecting the
main results presented below. Rather, they would lead to smaller
coupling constants necessary to reach a specific critical temperature
$T_c$. For example, we expect that taking into account all plasmon
acoustic plasmon branches would lead to higher critical temperatures (or
correspondingly to smaller Coulomb coupling constants for a given
$T_c$) since they
contribute to the attractive pairing interaction, though with lesser
weight.\cite{bill99,morawitz93}

As discussed in Sec.~\ref{ss:plasmons} the excitation spectrum of
a layered electron gas (see Fig.~\ref{Fig:pldisp}) allows us to divide
the $(q_\parallel,\omega)$-space into two main regions. The first
(region I) contains the additional collective excitations discussed
here and this region should be considered exactly in the equations
above. On the other hand, region II contains no such additional
plasmon features and the equations have a form analog to the 3D case
in this area. Therefore, we make a further step by dividing the
Coulomb part, Eq.~(\ref{gammactilde}), into two parts corresponding to
the two regions of Fig.~\ref{Fig:pldisp}:
\begin{eqnarray}
\label{twopartsGamma}
\Gamma_c(q_z,n-m) = \Gamma^{\rm I}_c(q_z,n-m) + \Gamma^{\rm
  II}_c(q_z,n-m)
=\hspace{1cm}&&\\
\hspace*{1.5cm}\frac{1}{\pi} \left\{\int_0^{\tilde{q}_c} +
  \int_{\tilde{q}_c}^1\right\}
\frac{\rm{d}\tilde{q}}{\sqrt{1-\tilde{q}^2}}\,\,
\frac{R(\tilde{q},q_z)/\tilde{q}}{\epsilon(\tilde{q},q_z,n-m)},\nonumber
\end{eqnarray}
where $q_c = \min\{2k_F, |\omega_n-\omega_m|/\hbar v_F\}$ and
$\tilde{q}_c \equiv q_c/2k_F$. The part $\Gamma_c^{\rm I}$ will be
considered exactly, in
particular with respect to the frequency dependence that has the
distinctive features of layered conductors. The part $\Gamma_c^{\rm II}$,
on the other hand, is reduced to an effective constant Coulomb
pseudopotential $\mu^\star\theta(\omega_n - \Omega_c)$ with a standard
cutoff given by $\Omega_c\simeq 10\times \Omega$ ($\Omega$ is the
characteristic phonon energy).
This treatment of region II calls for a comment. As was
  mentioned earlier we consider usual phonon-mediated
  superconductivity (Eliashberg equations) as a starting point of our
  analysis. Accordingly, the electron-phonon coupling constant and the
  static term $\mu^\star$ are treated in the conventional way as
  parameters to be determined from experimental data. Thus, the static
  term $\mu^\star$ is here a phenomenological parameter. We focus on
  the term
  describing the contribution of the dynamic screening. This part
  (present in region I) will be evaluated explicitely for different
  systems with the use of normal-state parameters (as, e.g., $v_F$ or
$\epsilon_M$; see below). Note
  that for the pure plasmon mechanism both static and dynamic terms
  were calculated in Ref.~\onlinecite{takada1} (in 3D). This step was
  crucial since its value was directly related to the
  criterion for the appearence of superconductivity. For the
  phonon-plasmon mechanism, on the other hand, we assume that the
  phonons are sufficient for the occurence of the superconducting
  state, which allows to use the conventional approach for the static
  term. Naturally, it would be of great interest to
calculate the static term. Such a full self-consistent calculation
including also real band structures will be carried out elsewhere.

As shown in appendix \ref{a:integ} (see also Ref.~\onlinecite{owen}),
the above equations can be mapped onto an eigenvalue problem written
in tensorial form
\begin{eqnarray}
\label{eigenvaleqtensor}
\underline{\underline{\bf K}}\,\, \underline{\Phi} = \eta\,
\underline{\Phi},
\end{eqnarray}
where $\left(\underline{\Phi}\right)_{n,n_z} = \Phi_n(n_z) \equiv
\Delta_n(n_z)/\sqrt{2n+1}$ and
$\underline{\underline{\bf K}}$ is given by Eq.~(\ref{Knm}) in
appendix \ref{a:integ}. Eq.~(\ref{eigenvaleqtensor}) is written
explicitely in Eq.~(\ref{eigenvaleq}). Note that an
artificial eigenvalue $\eta$ has been introduced in
Eq.~(\ref{eigenvaleqtensor}). $T_c$ is reached when $\eta$ is
one. Since all eigenvalues satisfy the inequality $\eta \leq 1$ we
only need studying the highest of them. Furthermore, the solution of
these equations give also the renormalized order parameter $\Phi_n$
near $T_c$. We can thus analyse how this function is affected by the
contribution of low-energy plasmons to the pairing interaction. A
typical function $\Phi_n$ is shown in
Fig.~\ref{Fig:PHIPlnitride} and will be discussed in the next
section. Using the eigenvalue equation (\ref{eigenvaleqtensor}) [or
(\ref{eigenvaleq}) in appendix \ref{a:integ}] we apply the theory to
various layered superconductors and calculate their $T_c$.

\subsection{Intercalated metal halide nitrides}\label{ss:nitrides}

The first class of materials we consider is the family of layered
metal-intercalated halide nitrides. We give
special attention to this family because low-energy plasmons not only
contribute to the pairing but, in fact, play the key role for
the superconducting state, as we show below. We believe that this is
the first observed system where the superconducting state of the
electrons is essentially self-supported, that is, where the pairing is
provided by collective excitations of the same carriers as those forming
pairs.

This family of novel superconductors has been discovered recently and
studied in detail in
Refs.~\onlinecite{yamanaka98,kawaji97,shamoto98,shamoto99,adelmann99,tou01,cros03,yamanaka01,tou03,yokoya01,weht99,hase99}.
As is known (see, e.g., Ref.~\onlinecite{yamanaka98}) the
intercalation of alkali atoms and organic molecules into the parent
compound (Zr,Hf)NCl leads to a superconductor with rather high
critical temperature ($T_c\sim 25$K). Based on
experimental\cite{yamanaka98,kawaji97,shamoto98,shamoto99,adelmann99,tou01,yamanaka01,tou03,yokoya01} studies and band structure
calculations\cite{weht99} it was concluded that electron-phonon
mediated pairing is insufficient to explain the observed $T_c$, since
the electron-phonon coupling constant appears to be too
small. Note also that a small nitrogen isotope effect of $T_c$ has
been observed.\cite{tou03} In addition, the compounds do not contain
any magnetic ions
and no sign of magnetism has been found in band
  structure calculations. This excludes a magnetic pairing
mechanism. Finally, normal-state properties of the materials studied
intensively in Refs.~\onlinecite{weht99,hase99,tou01} can be described
by Fermi liquid theory, so that there is no evidence for the presence
of strong correlations.

We apply our approach to this novel layered system. We note from
Eqs.~(\ref{lamc},\ref{DeltaZ0xi}-\ref{twopartsGamma}), that
the evaluation of $T_c$ for a specific compound needs the knowledge of
following parameters: the interlayer distance $L$, the band mass $m_b$
and Fermi velocity $v_F$, and the
dielectric constant of the spacers $\epsilon_M$. In addition, the
evaluation of the phonon contribution to the pairing requires the
knowledge of the characteristic phonon frequency $\Omega$, the
electron-phonon coupling constant $\lambda$ and the Coulomb
pseudopotential $\mu^\star$.

Specifically, we consider Li$_{0.48}$(THF)$_y$HfNCl
(THF=tetrahydrofurane) as an example, since the largest amount of
information necessary for the determination of $T_c$ is available for
this material, both from experiment and band structure
calculations. According to Refs.~\onlinecite{adelmann99,tou01,cros03}
the interlayer distance $L$ and the characteristic phonon frequency
$\Omega$ are equal to:
$L = 18.72$\AA\ and $\Omega=60$meV. The values of the band mass and
Fermi energy have been evaluated from band structure calculations,
Ref.~\onlinecite{weht99}. Accordingly, $m_b =0.6 m_e$, where $m_e$ is
the free electron mass and $\varepsilon_F\simeq 1$eV. For $\epsilon_M$
we have chosen the reasonable value $\epsilon_M = 1.75$. It
follows that $\lambda_c \simeq 0.8$ and, correspondingly, using
Eq.~(\ref{lamc}), $r_s \simeq 2$ (i.e.~close to the high-density
limit). The value
of the electron-phonon coupling constant can be estimated from the
knowledge of the electron specific heat constant $\gamma$ and the band
density of states $N(0)$. Indeed, the electron-phonon interaction
renormalizes $\gamma$ as $\gamma = \gamma_b(1 + \lambda)$, where
$\gamma_b = 2\pi^2 N(0)/3$ is the free electron Sommerfeld
constant. The value of $\gamma$ was estimated in
Ref.~\onlinecite{tou01} to be $\gamma \simeq 1.1$ mJ/molK$^2$, whereas
band structure calculations\cite{weht99} give $N(0) \simeq 0.74$
eV$^{-1}$. Thus, $\lambda \simeq 0.25$. Setting $\mu^\star = 0.1$ and
using Eqs.~(\ref{DeltaZ0xi}-\ref{twopartsGamma}), we obtain $T_c \simeq
24.5$K. The calculated $T_c$ is very close to the
observed value $T_c^{exp}=25.5$K. The essential point to note is that
in absence of the plasmon contribution we obtain $T_c^{phonon} \ll
1$K(!). This demonstrates that, indeed, the low-energy plasmon
contribution plays a key role for superconductivity in
metallochloronitrides.

It would be of great interest to carry out specific tunneling
(cf.~Ref.~\onlinecite{mcmillanwolf}) and optical measurements on this
material. We expect that tunneling experiments, similarly to heat
capacity data (see above), will provide the value $\lambda\simeq
0.25$, and optical measurements will lead to $\epsilon_M \simeq 1.75$.

\unitlength1cm
\begin{center}
\begin{figure}[htp]
  \includegraphics[width=7cm]{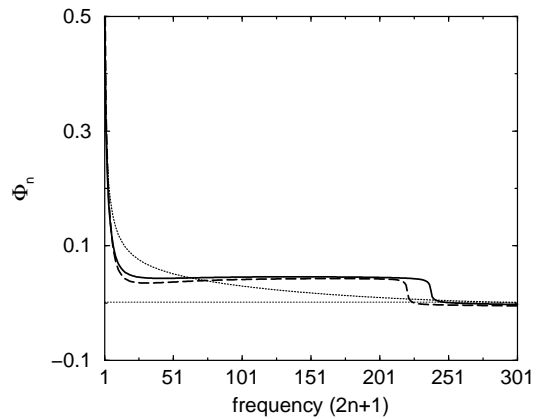}
\caption{\label{Fig:PHIPlnitride}
Normalized order parameter $\Phi_n=\Delta_n/\sqrt{2n+1}$ as a function
of Matsubara frequency $\omega_n = (2n+1)\pi T$. The solid and dashed
lines obtained for $\lambda=1$ and 1.5 respectively, represent the
order parameter in the presence of the plasmon contribution. The
dotted line was obtained in absence of the plasmon contribution
($\lambda_c=0$, $\lambda=1$). Note the presence of an additional
structure (step) when including the pairing due to low-energy plasmons
(solid and dashed lines). $\mu^\star=0.1$.}
\end{figure}
\end{center}
As mentioned earlier, by solving Eq.~(\ref{DeltaZ0xi}) we not only
obtain $T_c$, but we also get the superconducting order parameter
$\Phi_n=\Delta_n/\sqrt{2n+1}$. It is interesting to see how
$\Phi_n$ is affected by the additional pairing arising from the
presence of acoustic plasmons (Fig.~\ref{Fig:PHIPlnitride}). Note that
the qualitative form of the order parameter is
the same for all classes of materials discussed in the present paper.
In the absence of the plasmon contribution, the order parameter is a
rapidly decreasing monotonic function of the Matsubara frequency
(dotted line of Fig.~\ref{Fig:PHIPlnitride}). The effect of plasmons
reveals itself as an additional ``step'' in
$\Phi_n$ at
intermediate frequencies, as examplified by the solid and dashed lines on
Fig.~\ref{Fig:PHIPlnitride}. It is this positive part of the order
parameter due to the pairing induced by the low-energy collective
modes that is responsible for the enhancement of the value of
$T_c$. We observe 
that the order parameter remains positive over a frequency range
also determined by the value of $\lambda$ and $\mu^\star$. Whereas the
frequency range over which $\Phi_n$ remains positive shrinks with
increasing $\lambda$ (compare solid and dashed lines on
Fig.~\ref{Fig:PHIPlnitride}) it increases with increasing $\mu^\star$
(not shown on the figure). This apparently counter-intuitive behaviour
is easily understood by the fact that an order parameter extending to
higher frequencies will pick up more and more repulsive components of
the pairing interaction. The ``shorter'' the step in frequency of the
plasmon-induced structure, the smaller is the repulsive contribution
of the effective interaction kernel and, consequently, the higher is
$T_c$.

Concluding this section, we emphasize that the dynamic screening of
the Coulomb interaction (the contribution from low-energy electronic
collective modes) is essential for the understanding of the
superconducting state in intercalated layered metal halide nitrides.

\subsection{Layered organic superconductors}\label{ss:organics}

Organic superconductors were predicted in Ref.~\onlinecite{little64} and
discovered in Ref.~\onlinecite{jerome80}.
In this section we apply the theory to the class of layered organic
superconductors (see, e.g.,
Ref.~\onlinecite{jacobsen,tajima,kino,ishiguro,singleton02} and
references therein). As an example, we focus on
$\kappa-$(ET)$_2$Cu(NCS)$_2$ (ET=BEDT-TTF is a short notation for
bisethylenedithiotetrathiofuvalene). The basic structural building
blocks of these materials are large, elongated ET planar molecules
stretching along the $c$-axis forming the conducting
layer.\cite{urayama88} These
thick conducting layers are separated by thin insulating spacers of
planar NCS-molecules extending in the $ab$-plane. The NCS counter-ions
take one charge per two ET-molecules leaving the ET HOMO partially
unfilled. The HOMO are $\pi$-holes
delocalized over the large organic molecule and form the hole
conduction bands.

Layered organic conductors have highly anisotropic transport
properties. Typically, the ratio of in- to out-of-plane conductivity
is at least of the order $\sigma_{\parallel}/\sigma_{\perp} \sim
10^4$.\cite{ishiguro,mielke01} Band structure calculations\cite{xu95}
confirm the presence of quasi-two-dimensional bands. We emphasize once
more, however, that
only electronic transport properties are quasi-2D. As discussed in the
previous sections, the Coulomb interaction is important in all three
dimensions. In particular, incomplete screening
between layers implies that carriers from differents layers interact
with each other, leading to the low-energy electronic collective modes
discussed here. As we show in the following, this aspect is important
for understanding the relatively high value of the critical
temperatures observed in these materials.

Superconductivity has been observed for temperatures $T< T_c\simeq
10.4$K. Recent studies have shown the importance of electron-phonon
interaction for the pairing
mechanism.\cite{carlson,kini,pintchovius,sugai93,girlando02,varelogiannis02}
For example, isotope effect studies of the superconducting $T_c$ by
isotope substitution of C and S atoms on
the ET-molecules have singled out the effect of {\it intra}molecular
vibrations for the superconducting pairing \cite{carlson,kini}
A shift of phonon frequency caused by the superconducting transition
has also been observed with inelastic neutron
scattering.\cite{pintchovius} This shift indicates that the coupling
to {\it inter}molecular acoustic phonons contributes to
superconductivity. Further work supporting the importance of
electron-phonon interaction for superconductivity are given in
Refs.~\onlinecite{sugai93,girlando02,varelogiannis02}. Therefore, it
is interesting to apply our phonon-plasmon model to this class of
materials and study the effect of acoustic plasmons on the
superconducting $T_c$.

As it appears that both inter- and intramolecular vibrational modes
are of importance to superconductivity we modelize the phonon kernel
in Eq.~(\ref{gamma}) by a two-peak function
\begin{eqnarray}
\label{gammaphET}
\Gamma_{ph}(\omega_n-\omega_m) &=&\\
&&\hspace*{-2cm}\lambda\left[
w_1 \frac{\Omega_1^2}{(\omega_n - \omega_m)^2 + \Omega_1^2}
+ w_2 \frac{\Omega_2^2}{(\omega_n - \omega_m)^2+ \Omega_2^2} \right].
\nonumber
\end{eqnarray}
The lower frequency mode $\Omega_1 = 5$meV corresponds to libration
and intermolecular modes.\cite{ishiguro,pintchovius} The higher
frequency peak is located at the frequency $\Omega_2$, and was calculated
for the ET-intramolecular vibrations: $\Omega_2 =
10$meV.\cite{girlando02,elsinger00} Given the number of modes present
near each peak and their possible coupling to the electrons, we set
$w_1=0.75$ and $w_2=0.25$. The coupling constants to each set of modes
is then defined as $\lambda_j = \lambda w_j$ ($j=1,2$).

To calculate the value of $T_c$, we need to know the value of
the band mass $m_b$, the interlayer distance $L$, the Fermi energy
$\varepsilon_F$ and the dielectric constant of the spacers
$\epsilon_M$ (normal state parameters), as well as the value of the
electron-phonon coupling constant $\lambda$ and the Coulomb
pseudopotential $\mu^\star$. From band structure calculations
we have $m_b = 1.72 m_e$.\cite{xu95} The structure determination
gives $L= 16.2$\AA.\cite{urayama88} The average value of the
Fermi wave-vector obtained from Shubnikov-de Haas measurements is
$k_F\simeq 2.6\times 10^7$ cm$^{-1}$ 
(Ref.~\onlinecite{mielke01}). Inserting these values in $\varepsilon_F
\equiv \hbar^2 k_F^2/2m_b$ we obtain $\varepsilon_F \simeq
0.17$eV. Note that this is exactly the value obtained from band
structure calculations.\cite{xu95} Finally, we extract the value of
$\epsilon_M$ from optical reflectance measurements.\cite{ugawa88}
Using Eqs.~(\ref{epsilonomega}-\ref{epsilonM}) and the data of Ugawa
{\it et al.}\cite{ugawa88} we obtain $\epsilon_M =6.5$ (appendix
\ref{a:spacer}). Note that the ionic screening of the Coulomb
interaction is more efficient in organics than in
metallochloronitrides (previous section). One reason for this
difference is given by the fact that in organics the thick conducting
slabs are made of large molecules, whereas in the
metallochloronitrides conducting sheets are thin and made of
covalently bond atoms. The polarizability of the molecules implies
better ionic screening of the Coulomb interaction and, therefore, a
larger value of $\epsilon_M$. These parameters lead to $\lambda_c \simeq
0.9$ and thus $r_s\simeq 2.5$.

\begin{center}
\begin{figure}[htp]
\includegraphics[width=6.5cm]{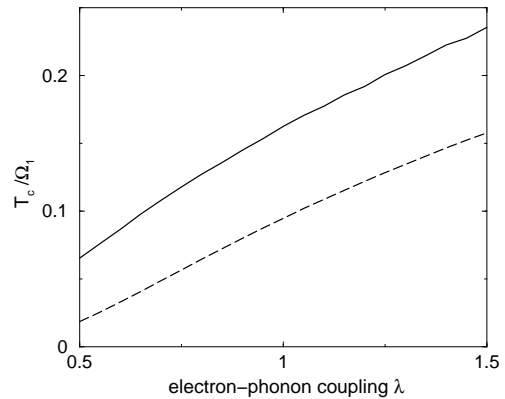}
\caption{\label{Fig:TclamET}
$T_c(\lambda)$ where $\lambda$ is defined in
Eq.~(\ref{gammaphET}). $T_c$ is normalized to the lowest phonon energy
$\Omega_1$. The upper (lower) curve is obtained in the presence
(absence) of the acoustic plasmon contribution. $\mu^\star=0.1$.}
\end{figure}
\end{center}

The exact value of the electron-phonon coupling constant $\lambda$ is
unknown at present. Estimates range $\lambda$ between 0.5 and
1.5.\cite{girlando02,elsinger00,ishiguro} Consequently, we
present results for this range of values in Fig.~\ref{Fig:TclamET} (we
have chosen $\mu^\star = 0.1$ and the cutoff at $\Omega_c =
10 \times \Omega_2 = 0.1$eV).
Using these parameters we calculate $T_c$ from
Eqs.~(\ref{DeltaZ0xi},\ref{gammabar}) (see
Fig.~\ref{Fig:TclamET}).
\unitlength1cm
The result shows that the increase of $T_c$ in the presence of low-energy 
electronic collective modes is substantial.
We can quantify this enhancement of $T_c$ for the specific example
studied, the $\kappa-$(ET)$_2$Cu(NCS)$_2$ compound. According to our calculation (see
Fig.~\ref{Fig:TclamET}) the experimentally observed value $T_c= 10.4$
K is obtained for $\lambda \simeq 1$, implying a coupling to
the low- and high-energy phonon modes of $\lambda_1 = 0.83$ and
$\lambda_2 = 0.28$, respectively. Thus, $\kappa-$(ET)$_2$Cu(NCS)$_2$
is an intermediate coupling superconductor. In the absence of
acoustic plasmons we obtain $T_c^{phonon} = 6.3$ K for this
$\lambda$. Thus, in the present case
$40\%$ of the value of $T_c$ is due to the pairing of electrons via
the exchange of acoustic plasmons. These calculations lead us to
conclude that the contribution of low-energy electronic collective
modes to the pairing is significant in organic superconductors (though
not dominant as in the case of metallohalidenitrides,
Sec.~\ref{ss:nitrides}).

\subsection{High-temperature oxides}\label{ss:HTSC}

In this section we discuss superconductivity in the cuprates within
our phonon-plasmon model. We analyze one specific material,
La$_{1.85}$Sr$_{0.15}$CuO$_4$, for which most parameters have been
determined. The normal state parameters are: the interlayer distance
$L=6.5$\AA, the Fermi wave-vector $k_F=3.5\times 10^7$ cm$^{-1}$, the
characteristic phonon frequency $\Omega \simeq 15$ meV, the dielectric
constant $\epsilon_M \simeq 5-10$.\cite{bozovic,falter02b}
Therefore, $\lambda_c \simeq 1$ and $r_s\simeq 2.8$. The
effective mass $m^\star$ and electron-phonon coupling constant
$\lambda$ were determined by Wolf and one of the authors from heat
capacity measurements.\cite{wolf90,kresinwolf} The obtained values are
$\lambda = 2$ and $m^\star \approx 5 m_b$. From the relation $m^\star =
(1+\lambda)m_b$ we then obtain $m_b\simeq 1.7$. Finally, the
Coulomb pseudopotential is taken to be $\mu^\star \simeq 0.1$.

The solution of Eqs.~(\ref{Delta0xi},\ref{Z0xi},\ref{gammabar}) with
use of the aforementioned parameters lead to $T_c = 36.5$K which is
close to the experimental value $T_c^{exp} \simeq 38$ K. It is
essential to note that in the absence of the screened Coulomb
interaction we would obtain $T_c^{ph} = 30$ K. Thus, about $20\%$ of
the observed value of $T_c$ is due to ``acoustic'' plasmons.\\
For thin films the stiffness of the lattice usually increases, leading
to a higher value of the characteristic phonon frequency
$\Omega$. Assuming $\Omega=20$ meV, we obtain $T_c = 49$K which is
close to the experimental value $T_c^{exp} = 45$ K observed, e.g.~in
Ref.~\onlinecite{kresinwolf}. This value is indeed higher than the one
of bulk samples. Interestingly, we obtain $T_c^{ph} = 40$ K, so that
the increase of $T_c$ induced by ``acoustic'' plasmons is again of
order of $20\%$.

Thus, the dynamically screened interlayer Coulomb interaction is
important for superconductivity in the cuprates. Note that a proper
account of the Coulomb interaction screening is not only of importance
for superconductivity in these materials, but also for a proper
description of normal state properties such as lattice
dynamics.\cite{falter02a,falter02b}

\section{Discussion and Conclusions}\label{s:conclusions}

The purpose of the present article was to study the impact of layering
on superocnductivity. Particular
emphasis was set on the {\it dynamically screened} Coulomb
interaction. Layered materials have distinctive {\it low-energy}
electronic collective excitations that provide exchange bosons for the
pairing between electrons. We have shown that these acoustic plasmons
lead to an enhancement of the superconducting $T_c$. We have applied
the theory to three classes of layered superconductors:
alkali-intercalated halide nitrides, organic and high-temperature
superconductors.

Within our phonon-plasmon model we observe an increasing influence of the
electronic pairing mechanism for the three classes of layered
superconductors considered. In metal-intercalated halide
nitrides the contribution arising from low-energy electronic
collective modes is dominant. These materials are thus unique in the
sense that an electronic pairing mechanism is at the origin of
superconductivity: the exchange bosons are made of the same particles
(the electrons) than those who bind into pairs below $T_c$. In
the case of organic layered materials, the electronic and phononic
energies, as well as the structure of the conducting layers and
insulating spacers, leads to a situation where the contribution of
phonons and acoustic plasmons is of the same order. Finally, in the
case of high-temperature superconductors, the contribution of
low-energy plasmons is significant but not dominant. Within our model
the phonon contribution is still largest.

There are other classes of layered superconductors that have not been
considered in the present article. Among them, the most prominent is
that of dichalcogenides. We believe that some
experimental observations\cite{geballe} are related to the phenomenon
discussed in this article. However, we also note that many of the
systems belonging to this class of materials exhibit charge density wave
instabilities. This both obscures and changes the contribution of
acoustic plasmons to superconductivity and will be discussed
elsewhere.

Another interesting system is the CoO$_2$-based layered compound
studied recently in Ref.~\onlinecite{takada03}. We point out that the
system becomes superconducting only for relatively large interlayer
distance. This is consistent with the present theory and the material
deserves further study.

An essential conclusion of the present work is that the physics of
layered (super)conductors cannot be reduced to the study of one
conducting layer (or the layers belonging to one unit cell as in some
high-temperature superconductors). Such simplification
relies on the observation of ``quasi-two-dimensional''
transport. However, it misses to account for the screening
properties of the electron-electron Coulomb interaction (and of the
electron-phonon interaction as well; see
Refs.~\onlinecite{bill97,falter02a}). As we
discussed in the paper, the screening is very different in layered
materials as compared to 2D and 3D isotropic
metals. We believe that the particular screening properties are
essential for the behaviour of layered (super)conductors. How large
the effect of
screened Coulomb interaction is, depends very much on the specific
features of the material. For example, the covalency within the
conducting layers and the structure of the spacers, the presence of
van der Waals gaps
will determine its contribution both to normal and superconducting
state properties. Therefore, the study of screening properties in
layered conductors is a promising direction to better understand the
similarities and differences between different classes of materials
and serve as a bridge in the study of properties of 2D and 3D systems.

\section{Acknowledgements}
We thank I.~Bozovic, D.~van der Marel and S.A.~Wolf for fruitful
discussions. H.M.~thanks P.~Reineker and the Institute of
Theoretical Physics of Ulm University for discussions and
hospitality. A.B.~thanks the Max Planck Institute for the Physics of
Complex Systems, Dresden, for hospitality.


\vspace*{1cm}
\begin{appendix}
\section{Coulomb potential for layered systems}
\label{a:Coulomb}
Using cylindrical coordinates ${\bf r} = ({\bf r}_{\parallel},z)$
(where $z$ is perpendicular to the layers) the Fourier transform of
the 3D Coulomb potential $V_c({\bf r}) = e^2/\epsilon_M |{\bf r}|$ is
given in layered structures by\cite{fetter}
\begin{eqnarray}
V_c({\bf q}) &=& \frac{1}{N_z}\sum_{n} e^{\mathrm{-i}q_znL}
\int \mathrm{d{\bf r}}_{\parallel}e^{\mathrm{-i}{\bf
    q}_{\parallel}{\bf r}_{\parallel}}
\frac{e^2\epsilon_M^{-1}}{{\bf r}_{\parallel}^2 + (nL)^2}\nonumber\\
\label{Vcz}
&=& \frac{1}{N_z}\sum_{n} e^{\mathrm{-i}q_znL}\,
\frac{2\pi e^2}{\epsilon_Mq_{\parallel}}e^{-q_{\parallel}nL},
\end{eqnarray}
where we have taken into account the fact that the charges are located
in the conducting sheets, and thus $z=nL$ where $L$ is the interlayer
spacing and $n$ indexes the layers. Note that the second line of
Eq.~(\ref{Vcz}) shows how the Coulomb interaction is exponentially
decaying (in real space) along the direction perpendicular to the
layers, the decay being determined by $q_\parallel L$. Performing the
sum in Eq.~(\ref{Vcz}) we obtain Eq.~(\ref{Vcq}).

Note that the detailed structure of the spacers separating the
conducting sheets is not considered in the present model. We thus have 
included the screening resulting from polarization effect of the
spacers via the dielectric constant $\epsilon_M$. The dielectric
function in the denominator of Eq.~(\ref{gammac}) thus accounts for
the screening induced by the charge-carriers of the conduction bands
only.

\section{Polarizability}
\label{a:pol}

The RPA polarizability of a single conducting sheet, Eq.~(\ref{pol}),
is written in polar coordinates ${\bf k}_\parallel = (k,\varphi)$
\begin{eqnarray}
\label{PiN}
\Pi({\bf q}_\parallel,i\omega_n) &=& \frac{2}{(2\pi)^2}
\int_0^\infty {\rm d}k\, k\,f_k\, I_\varphi(k, {\bf q}_\parallel,
\omega_n),\\
\label{Iphi}
I_{\varphi}(k,{\bf q}_\parallel,i\omega_n) = &&\\
&&\hspace*{-3cm}
\int_0^{2\pi} {\rm d}\varphi
\left\{ \frac{1}{i\omega_n + \xi_{{\bf k}_\parallel} + \xi_{{\bf
        k}_\parallel + {\bf q}_\parallel}}
- \frac{1}{i\omega_n + \xi_{{\bf k}_\parallel} - \xi_{{\bf
      k}_\parallel + {\bf q}_\parallel}} \right\}.\nonumber
\end{eqnarray}
For $T=0$ the integral over ${\bf k}_\parallel$ can be calculated
analytically, leading to the result first derived in
Ref.~\onlinecite{stern}. At finite temperature there is no simple
analytical form. However, to reduce the amount of numerical work in
solving Eqs.~(\ref{Delta0xi},\ref{Z0xi}) for $T_c$, we calculate the
angle integral analytically. Using the transformation $z =
\exp(i\varphi)$ we integrate $I_\varphi$ over z in the complex plane
to obtain
\begin{eqnarray}
I_\varphi &=& -
\frac{\sqrt{2}\pi}{q_\parallel}\,\frac{u_+}{W}
\left[\delta\right(\tilde{k}\left) + \theta_1 -
  \theta_2 \right],
\end{eqnarray}
with $u_{\pm} = \sqrt{A \pm W}$, $W = \sqrt{A^2 + B^2}$, $A =
4\tilde{k}^2\tilde{q}^2(\zeta_1^2 - \zeta_2^2 -1)$,
$B=4\tilde{k}^2\tilde{q}^2\zeta_1\zeta_2$, and
\begin{eqnarray}
\label{gamma12}
\zeta = \zeta_1 + i \zeta_2
\equiv \frac{\tilde{q}}{2\tilde{k}} +
i\frac{\tilde{\omega}_n}{4\tilde{k}\tilde{q}}.
\end{eqnarray}
As in the main text we normalize all wave-vectors to $\tilde{X} \equiv
X/2k_F$, $X = q,k$.
$\theta_j$
($j=1,2$) are defined in terms of Heavyside functions as
$\theta_j = \theta(1 - |z_j|^2$), $z_j = x_j + i y_j$ with
\begin{eqnarray}
x_1 = |u_+|-\zeta_1, & y_1 = |u_-|-\zeta_2,\\
x_2 = -|u_+|-\zeta_1,& y_2 = -|u_-|-\zeta_2,
\end{eqnarray}
for $B\geq 0$ whereas $y_1$ and $y_2$ are interchanged for $B<
0$. Inserting this expression into Eq.~(\ref{PiN}) above, we obtain
the following compact form for the polarizability
\begin{eqnarray}
\label{PiNInt}
\Pi(\tilde{q},\omega_n) &=& - N(0)\frac{\sqrt{2}}{\tilde{q}^2}
\int_0^\infty {\rm d}A f_A \frac{\partial u_+}{\partial A} 
\left[\delta\right(\tilde{k}\left) + \theta_1 -
  \theta_2 \right].
\end{eqnarray}
This expression of the polarizability has been used to calculate the
dielectric function that appears in Eqs.~(\ref{Delta0xi},\ref{Z0xi})
and is depicted in Fig.~\ref{Fig:polLEG}. Note that
\begin{eqnarray}\label{polq01}
\lim_{q\to 0} \Pi(q,\omega_n-\omega_m) = - 2 N(0) f_{\bf k_F}
\delta_{\omega_n,\omega_m}.
\end{eqnarray}

\section{Equations for the order parameter and the renormalization
  function}\label{a:integ}

We start with Eq.~(\ref{phiZ0}) and, as mentioned in
Sec.~\ref{ss:numerics}, assume isotropy of the bands within the
planes. Thus, $\Delta$ and $Z$ depend only on the norm of ${\bf
  k}_\parallel$ (and on $k_z$). In this case, it is possible to
calculate one of the integrals over ${\bf k'}_\parallel$
analytically. To this
aim, we transform the 2D in-plane integration in a way analogous to
the 3D cases generally studied, namely introducing polar coordinates
$\rm{d}^2{\bf k} = k'\rm{d}k'\,\rm{d}\varphi$. With $k'\rm{d}k' = 2\pi
N(0)\,\rm{d}\xi$ and $q^2 = |{\bf k}'-{\bf k}|^2 = k^2 + {k'}^2 -
2k'k\cos(\varphi)$ the integral over ${\bf k}_\parallel$ is
transformed into an integral over energy $\xi$ and angle
$\varphi$. Using the fact that the resulting energy integrand of
Eq.~(\ref{phiZ0}) is falling off as $\xi^{-2}$, the main contribution
to this integral will come  from $\xi/\varepsilon_F \ll 1$ and we obtain
\begin{eqnarray}
\label{kpphiq}
\rm{d}^2{\bf k'}&=& 2\pi N(0)\rm{d}\xi \rm{d}\varphi
\simeq  4\pi N(0) \rm{d}\xi \frac{\rm{d}\tilde{q}}{\sqrt{1-\tilde{q}^2}},
\end{eqnarray}
with $\tilde{q}< 1$.
The energy integral can then easily be performed. Gathering the different
terms and assuming that the electron-phonon coupling function
$g_\nu({\bf q}) = g_\nu(q_\parallel)$, in order to define $\lambda$ and
$\Gamma_c$ as in Eq.~(\ref{gammabarparts}), we obtain
Eq.~(\ref{DeltaZ0xi}). The latter equations have been obtained by
discretising $k_z = -\pi/L + 2\pi n_z/N_zL$ with
$n_z=1,\dots,N_z$. Note, that the angle-integration can in principle be
performed exactly, without need of the approximation,
Eq.~(\ref{kpphiq}). However, the difference with the present method is
minimal and we use the approximation above for simplicity.

To perform the numerical calculation we cast Eq.~(\ref{phiZ0}) or
(\ref{DeltaZ0xi}) into a matrix form. We first transform the summation
over $m=\dots -1,0,1,\dots$ to a sum over non-negative $m$
only. Equation (\ref{DeltaZ0xi}) then takes the form
\begin{widetext}
\begin{subequations}
\label{DeltaZ0xif}
\begin{eqnarray}
\label{Delta0xif}
\Delta_n(k_z) Z_n(k_z)&=&
\pi T \sum_{m\ge 0}\,\,
\frac{1}{N_z} \sum_{k_z'=-\pi}^\pi\,
\left\{\bar{\Gamma}(q_z,n-m) + \bar{\Gamma}(q_z,n+m+1)\right\}\,
\frac{\Delta_m(k_z')}{|\omega_m|},\\
\label{Z0xif}
Z_n(k_z) &=& 1 + \,\pi \frac{T}{\omega_n} \sum_{m=0}^{2n}
\frac{1}{N_z} \sum_{k_z'=-\pi}^\pi \bar{\Gamma}'(q_z,n-m).
\end{eqnarray}
\end{subequations}
\end{widetext}
The second equation has been simplified further, reducing the sum over
$m$ to the range $[0,2n]$. The kernel $\bar{\Gamma}'$ in
Eq.~(\ref{Z0xif}) now only contains frequency-dependent terms. All
frequency-independent terms vanished in the folding of the summation
over $m$.

Inserting Eq.~(\ref{Z0xif}) into (\ref{Delta0xif}), defining 
$\Phi_n(k_z)=\Delta_n(k_z)/\sqrt{2n+1}$ and
\begin{widetext}
\begin{eqnarray}
\label{Knm}
K_{nm}(q_z=k_z'-k_z) =
\frac{1}{\sqrt{(2n+1)(2m+1)}}
\left\{
  \bar{\Gamma}(q_z,n-m) + \bar{\Gamma}(q_z,n+m+1)
  - \delta_{n,m} \sum_{p=0}^{2n}\bar{\Gamma}'(q_z,n-p),
\right\}
\end{eqnarray}
\end{widetext}
we finally condense Eq.~(\ref{DeltaZ0xif}) to the matrix form
($q_z=k_z'-k_z$)
\begin{eqnarray}
\label{eigenvaleq}
\sum_{m\ge 0} 
\frac{1}{N_z} \sum_{n_z'=1}^{N_z}\,
K_{nm}(|n_z'-n_z|)\, \Phi_m(n_z')
&=&\\
&&\eta\,\Phi_n(n_z),\nonumber
\end{eqnarray}
$n_z=1,\dots,N_z$. This is the explicit form of Eq.~(\ref{eigenvaleqtensor}). Note that the
kernel $K_{nm}(q_z)$ depends on $n$ and $m$ separately and not only on
$n-m$. Furthermore, the kernel is even in $q_z$, $K_{nm}(q_z) =
K_{nm}(|q_z|)$. We have introduced the artificial eigenvalue
$\eta$ to map the problem onto an eigenvalue equation. $T_c$ is
obtained when $\eta=1$.

\section{Dielectric constant of the spacers $\epsilon_M$}\label{a:spacer}

The dielectric constant of the spacers $\epsilon_M$ can be extracted
from infrared or reflectivity data. We parametrize the dielectric
function obtained in these experiments by the Drude-Lorentz model
\begin{eqnarray}
\label{epsilonomega}
\epsilon(\omega) = \epsilon_\infty
+ \sum_j \frac{S_j \omega_j^2}{\omega_j^2 - \omega^2 - i \omega\Gamma_j}
+ \epsilon_{\rm fc},
\end{eqnarray}
where $\epsilon_{\rm fc}$ is the free carrier contribution to the
dielectric constant. The dielectric constant for the spacers is then
defined as
\begin{eqnarray}
\label{epsilonM}
\epsilon_M = \epsilon(\omega=0) - \epsilon_{\rm fc}
 = \epsilon_\infty + \sum_j S_j.
\end{eqnarray}

For the determination of the dielectric constant of the organic
material $\kappa-$(ET)$_2$Cu(NCS)$_2$ we use Ugawa {\it et al.}'s
reflectivity measurements.\cite{ugawa88} Their parametrization gives
$\epsilon_\infty = 3.2$ and
\begin{center}
\begin{math}
\begin{array}{lll}
\omega_1 = 0.16 & \omega_2 = 0.28 & \omega_3=0.47 \rm{eV},\\
\Omega_{p1} = 0.093 & \Omega_{p2} = 0.7 & \Omega_{p3}=0.44 \rm{eV}.
\end{array}
\end{math}
\end{center}
With the correspondance $\Omega_{pj}^2 \equiv S_j \omega_j^2$ we have
$S_1\simeq 0.762$, $S_2\simeq 1.581$, $S_3\simeq 0.968$. From these
data and Eq.~(\ref{epsilonM}) it follows that
\begin{eqnarray}
\label{epsilonMval}
\epsilon_M = 6.5
\end{eqnarray}

\end{appendix}

\end{document}